\documentclass[nonacm,manuscript,screen]{acmart}
\AtBeginDocument{%
  }

\begin{document}

\title{Functional Misalignment in  Human--AI Interactions on Digital Platforms}

\author{Kristina Lerman}
\email{krlerman@iu.edu}
\affiliation{%
  \institution{Indiana University}
  \city{Bloomington}
  \state{Indiana}
  \country{USA}
}

\renewcommand{\shortauthors}{Lerman}

\begin{abstract}
Algorithmic systems, particularly social media recommenders, have achieved remarkable success in predicting behavior. By optimizing for observable signals such as clicks, views, and engagement, these systems effectively capture user attention and guide interaction. Yet their widespread adoption has coincided with troubling outcomes, including rising mental health concerns, increasing polarization, and erosion of trust. This paper argues that these effects are consequences of a structural \emph{functional misalignment} between what algorithms optimize---predictable behavior---and the human goals these predictions are intended to serve. We propose that this misalignment arises through three mechanisms: (1) a bias toward modeling fast, reactive behavioral signals over reflective judgment, (2) feedback loops that couple user behavior with algorithmic learning, and (3) emergent collective dynamics that amplify these effects at scale. Together, these mechanisms explain how accurate individual-level predictions can produce adverse societal outcomes. We present functional misalignment as a unifying framework and outline a research agenda for studying and mitigating its effects in human–AI interaction systems.
\end{abstract}




\maketitle

\section{Introduction}

Advances in machine learning and large-scale computing have enabled algorithmic systems to predict and shape human behavior at an unprecendented scale. Social media recommender systems, in particular, excel at forecasting what users will click, watch, or share. TikTok is said to ``read your mind''~\citep{Smith2021TikTok}, predicting within seconds of a new user's first session, which videos they will watch.  
As a result, these systems have become highly effective at capturing attention and sustaining engagement at an industrial scale, leading users to spend substantial portions of their daily lives interacting with algorithmically curated feeds. For example, American teenagers now spend an average of eight hours per day on screen-based digital media, including approximately 1.5 hours on social media~\citep{rideout2021common}.

Algorithmic systems are now being blamed for a host of troubling societal and individual outcomes, from rising mental health concerns among adolescents, to growing political polarization, erosion of democratic norms and trust, and other forms of collective dysfunction. How can systems that are so effective at predicting human behavior produce outcomes that appear misaligned with social welfare?

This paper argues that these broad ills can be understood as an emergent property of  \emph{functional misalignment} within human-algorithm interactions. We define functional misalignment as a systematic mismatch between what algorithmic systems are optimized to do, namely, accurately predict observable behavior, and the underlying human goals these predictions are intended to serve, such as satisfying user preferences, values, needs, and well-being. In social media environments, this mismatch is particularly acute because the signals that are most easily observed and predicted---rapid, emotional, and reactive behaviors---are not necessarily those that reflect users' considered interests or long-term welfare.

We propose that functional misalignment arises through the interaction of three mechanisms. First, algorithmic systems optimize over observable behavioral signals that disproportionately reflect fast, heuristic-driven responses rather than deliberative, reflective judgment. Second, interactions between people and algorithms create feedback loops. Third, these dynamical processes result in emergent large-scale phenomena in social systems, not all of them desirable. 

Together, these mechanisms help explain why successful predictions at the individual level can coexist with adverse societal outcomes. They also suggest that many observed problems, such as rising polarization and growing youth mental health crisis, share a common structural origin in how algorithmic systems learn from and influence human behavior.

This paper makes three primary contributions. First, it describes functional misalignment as a unifying framework for understanding human--AI interaction at scale. Second, it integrates insights from cognitive science, machine learning and complex systems  to identify the mechanisms through which this misalignment emerges. Third, it articulates a research agenda for studying and mitigating these effects, including new directions for measurement, experimentation, and system design.

By clarifying the mechanisms underlying algorithmic misalignment, we aim to enable a more coherent and cumulative research program on human--AI interaction, one that can inform both technical design and policy interventions in large-scale digital systems.

\section{Background and Related Work}

The study of human--AI interaction draws on three traditions: cognitive science, which characterizes human judgment and its limitations; machine learning, which explains how algorithms learn from behavior; and complex systems theory, which analyzes the dynamics that emerge when these processes interact.

\subsection{Cognitive Science Perspective}

Human cognition evolved to manage complex social environments, including reputation tracking, threat detection, and coalition formation~\citep{li2018evolutionary}. This architecture relies heavily on cognitive heuristics---fast, automatic decision rules that trade accuracy for efficiency~\citep{tversky1974judgment}. Dual-process theories distinguish between a rapid, heuristics-driven System~1 and a slower, deliberative System~2 \citep{kahneman2011thinking}.

System~1 generates quick, intuitive responses that are often adaptive but systematically biased. For example, negatively-valenced information attracts greater attention and is more readily believed than positive information, because ignoring it can be catastrophic to individuals~\citep{baumeister2001bad,fessler2014negatively}. These biases produce predictable behavioral patterns. System~2 supports reasoning and reflection, but requires effort, and under cognitive load or emotional arousal,  individuals often to default to System~1 responses.

\subsection{Machine Learning Perspective}
\label{sec:machine-learning}
Recommender and ranking systems span a spectrum from simple aggregation of user signals (e.g., popularity-based ranking) to fully personalized models that predict individual behavior based on past interactions. Many systems also employ collaborative filtering, leveraging patterns across users to recommend new items~\citep{pedreschi2025human}. In all cases, these methods generate rankings that shape what users see.

Modern recommender systems are trained to optimize predictive accuracy over behavioral signals such as clicks, views, and engagement. This approach assumes that behavior is a useful proxy for what users want, need or value. While effective for personalization, this assumption is imperfect: engagement-based objectives do not necessarily align with broader notions of what users want~\citep{milli2021optimizing}. Behavioral signals, especially in high-volume settings, disproportionately reflect fast, affective System~1 processes, leading algorithms to learn patterns associated with reactive rather than reflective behavior.

\subsection{Complex Systems Perspective}
\label{sec:complex-systems}
Human--AI interaction can be understood as a complex system driven by feedback loops. In such systems, outputs influence future inputs, shaping its subsequent behavior. Positive feedback loops amplify small initial differences into large disparities, an effect known as the rich-get-richer effect or cumulative advantage~\citep{Merton1968}.

In algorithmic settings, these dynamics produce several effects: preference drift, reduced diversity, and path dependence \citep{carroll2021estimating,pedreschi2025human}. As we argue, feedback loops can also create instability and unpredictability, making long-term system behavior difficult to control.

\subsection{Relation to Existing Work}

\subsubsection{Evolutionary Mismatch and Cognitive Biases}

The evolutionary mismatch hypothesis explains why cognitive heuristics and biases are particularly potent in digital environments: mechanisms adapted to ancestral contexts can produce maladaptive outcomes in modern settings~\citep{li2018evolutionary,lim2024social}. Social media platforms exploit sensitivities related to status, belonging, and intergroup competition, making emotionally charged content especially engaging.

Empirical work shows that cognitive biases interact with algorithmic ranking. Position bias shapes attention and limits information diffusion~\citep{lerman2014leveraging,lerman2016future}. In crowdsourcing, ranking combined with biased decision-making can distort collective outcomes~\citep{burghardt2020origins}. As information load increases, reliance on heuristics intensifies, making behavior more predictable~\citep{burghardt2017myopia}. 

Despite these insights, cognitive and algorithmic perspectives are often studied separately. We argue that recommender systems actively amplify these biases by learning to trigger predictable System~1 responses.

\subsubsection{Algorithmic Amplification of Polarization}

A large literature documents how social media algorithms shape information environments and contribute to filter bubbles, polarization and emotional divides. Platform audits reveal systematic differences between algorithmic and chronological feeds \citep{bartley2021,huszar2022algorithmic}. Emotionally charged content spreads more widely on social media~\citep{brady2017emotion}, and algorithm-mediated learning reinforces such dynamics~\citep{brady2023algorithm}. Engagement optimization favors extreme and polarizing content~\citep{shin2024misinformation}.

While some work argues that cross-cutting exposure remains common~\citep{barbera2020}, polarization is now thought to arise through affective mechanisms rather than ideological isolation. Models show that out-group animosity alone can produce polarization \citep{nettasinghe2025out,nettasinghe2025group}, and empirical work links platform dynamics to partisan sorting~\citep{tornberg2022,chen2021neutral}.
%
While this perspective captures system dynamics, it remains focused largely on polarization and does not explain other societal phenomena.

\subsubsection{Algorithmic Misalignment}

A growing literature examines the gap between algorithmic objectives and human values.\citet{kleinberg2024inversion} formalize the \emph{inversion problem}: algorithms must infer latent mental states from behavioral signals that are imperfect proxies. \citet{milli2021optimizing} show that engagement is not a reliable measure of value, and subsequent work demonstrates that engagement optimization can amplify divisive content while reducing user satisfaction~\citep{milli2025engagement}.

While this literature identifies the existence of misalignment, it does not fully explain the mechanisms through which it arises. In particular, the interaction between cognitive biases, feedback loops, and algorithmic learning remains under-theorized.

\subsubsection{Synthesis and Gaps}

Across these strands, prior work identifies key elements: cognitive biases that shape behavior, feedback loops that amplify outcomes, and a gap between engagement and user welfare. However, these insights are typically studied in isolation. Existing work treats polarization, radicalization, and psychological harms as distinct phenomena, without a unified explanation.

This paper integrates these perspectives into a single framework. We argue that these outcomes are not isolated failures but emergent consequences of a structural mismatch between optimization over observable behavior and the requirements of human welfare. By identifying the mechanisms---mismatched learning objectives, feedback-driven dynamics, emergent phenomena---we provide a unified account of how human--AI interactions generate and sustain these effects.

\section{Functional Misalignment in Human--AI Interactions}

We define \emph{functional misalignment} in human--AI interactions as a structural mismatch between what algorithmic systems optimize and what people want, need or value. Modern recommender and ranking systems are trained to predict and maximize \emph{observable user behavior}---clicks, likes, shares, and dwell time---using these signals as proxies for preferences, values, identity, and well-being. 

This creates an \emph{epistemic gap} between what is measurable and what ultimately matters for human welfare. This misalignment is \emph{functional}, not intended: systems perform as designed, but optimize over signals that systematically diverge from the internal  dimensions of what people value.

\subsection{Mechanisms of Functional Misalignment}

\begin{figure}
    \centering
    \includegraphics[width=0.8\linewidth]{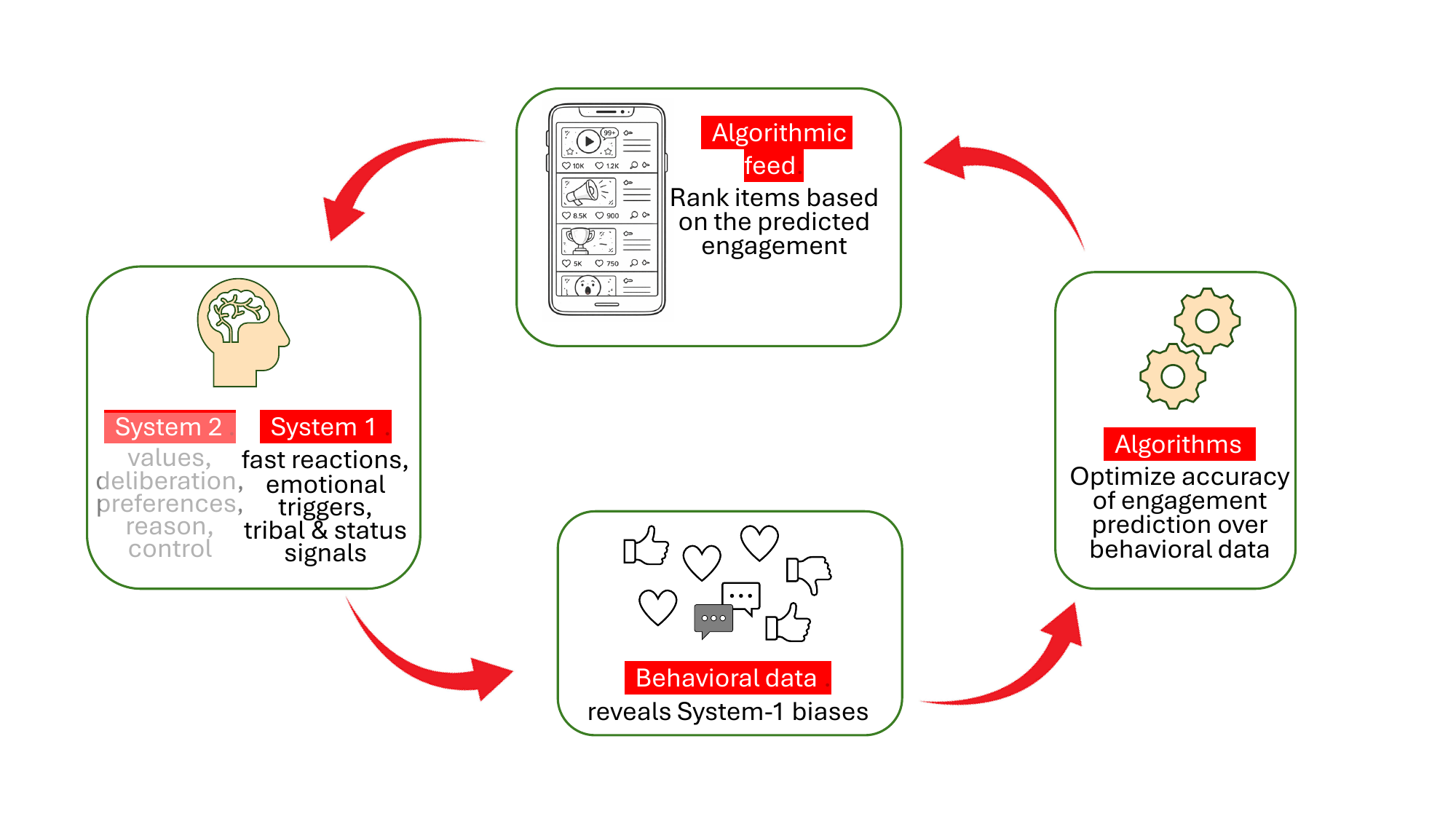}
    \caption{Functional misalignment framework: Social media algorithms trained to accurately predict user engagement learn to amplify System-1 biases (outrage, envy, status-seeking, in-group cues), creating feedback loops that systematically misalign platform outcomes from users’ reflective preferences and collective welfare.}
    \label{fig:diagram}
\end{figure}

We argue that functional misalignment arises through three interacting mechanisms: (1) observability and predictability bias, (2) feedback-loop dynamics, and (3) emergent collective effects. Figure~\ref{fig:diagram} illustrates these mechanisms.

\subsubsection{Observability Mismatch}

Algorithmic systems rely on behavioral data to infer user preferences, often treating observed actions as revealed preferences. However, behavior reflects a mixture of cognitive processes. System~1 responses---fast, affective, and heuristic-driven---are more immediate, frequent, and consistent across users than deliberative System~2 processes~\citep{kahneman2011thinking}. 

Because System~1 responses are easier to predict, especially under conditions of information overload~\citep{burghardt2017myopia}, optimization over behavioral data systematically privileges these signals. As a result, algorithms learn to model and optimize reactive behaviors, even when these diverge from users’ reflective preferences or long-term welfare~\citep{milli2025engagement}.

\subsubsection{Feedback-Loop Dynamics}
\label{sec:feedback}

Recommender systems operate within closed feedback loops: algorithmic outputs influence user behavior, which becomes new training data:
\begin{quote}
\begin{center}
Behavior $\rightarrow$ Data $\rightarrow$ Algorithm $\rightarrow$ Exposure $\rightarrow$ Behavior
\end{center} (see Fig.~\ref{fig:diagram}).    
\end{quote}
These feedback loops create a dynamical system in which small initial differences are amplified over time through self-reinforcing, rich-get-richer processes. Such feedback loops exhibit characteristic pathologies: they are inherently \textit{unpredictable} (small early fluctuations compound into large and irreversible disparities)~\citep{salganik2006experimental};  produce extreme \textit{inequality} (few items capture a disproportionate share of attention); and they erode \textit{diversity} (by homogenizing the content and perspectives)~\citep{pedreschi2025human}. Additionally, they generate systematic \textit{disparities}: feedback loops amplify even small initial biases---whether based on prestige, gender, or group membership---into large group-level inequalities \citep{nettasinghe2026emergence}.

\subsubsection{Emergent Collective Effects}

The third mechanism concerns the large-scale consequences of deploying feedback-driven systems across populations of millions. Because algorithms prioritize engagement, they disproportionately amplify content that triggers strong System~1 responses: anger, outrage, fear, envy, and tribal identification. The system learns to surface whatever most reliably elicits predictable reactions, regardless of broader social consequences.

At scale, these dynamics produce emergent phenomena: polarization, norm misperception, amplification of extreme viewpoints, intergroup hostility and unfairness. They resemble the behavior of complex systems more broadly, where local optimization produces unintended and sometimes harmful global effects.



\section{Pathological Outcomes of Functional Misalignment}

We highlight three domains where the harmful emergent effects of functional misalignment may be most evident.

\subsection{Political Polarization}
Political polarization has 
widened dramatically over the past two decades in the United States. Increasingly, polarization is manifested as emotional division between in-groups and perceived out-groups rather than substantive disagreement over policy. Political scientists conceptualize this as \textit{affective polarization}, comprising in-group favoritism on one side and out-group animosity on the other. In affectively polarized environments, political opponents are perceived as immoral or dangerous and differences become existential threats~\citep{iyengar2015fear}.

Conventional accounts attribute these trends to misinformation, coordinated inauthentic behavior, and echo chambers on social media~\citep{lewandowsky2012misinformation, nikolov2015measuring, cinelli2021echo, chen2021neutral}. But these can be better understood as symptoms than as root causes. Misinformation spreads more readily in affectively polarized populations where out-group distrust is already high; coordinated manipulation gains traction where emotional divisions have weakened social ties; malicious manipulation succeeds by exploiting preexisting grievances. Treating these downstream phenomena as the primary targets of intervention leaves the generative mechanism intact.

That mechanism, we argue, is algorithmic amplification of intergroup hostility. Polarization is not a static property of ideologically sorted populations but emerges from the interactions between emotionally-divided groups. Social media algorithms amplify anger and moral outrage directed at out-groups~\citep{brady2017emotion}, and users adapt their own expressive behavior toward content that receives high engagement, reinforcing animosity~\citep{brady2023algorithm}. The content surfaced by algorithmic feeds is therefore systematically skewed toward intergroup hostility, deepening emotional divisions of affective polarization.

The collective consequence of these micro-level emotional dynamics is societal ideological division. In affectively polarized societies, small perturbations---such as identity threats or elite signaling~\citep{remso2025climate,wendsjo2025male}---can create durable partisan polarization in attitudes and perceived norms~\citep{nettasinghe2025out} even in non-political domains~\citep{nettasinghe2025group}.
In this view, polarization is not simply ideological separation but an emergent property of intergroup hostility, sustained and amplified by algorithmic interactions.

\subsection{Mental Health and Collective Wellbeing}

A growing literature links social media use to adverse mental health outcomes, particularly among children and adolescents. Rising rates of anxiety, depression, and self-harm have been widely documented~\citep{twenge2017igen,haidt2024anxious}, with adolescent girls disproportionately affected~\citep{radhakrishnan2022pediatric}. Internal research by social media platforms suggests many users feel worse about themselves after use~\citep{olesen2025big}. While causal impacts of social media on mental health outcomes remain debated~\citep{orben2024mechanisms}, meta-analyses consistently find associations between increased use of digital platforms and poor mental health outcomes~\citep{teague2026digital}.

The proposed framework suggests a common mechanism linking mental health and polarization: algorithms amplify content that triggers predictable affective responses. Among adolescents, these responses often involve social comparison, envy, and status threat~\citep{choukas2022perfect}. Because such signals are immediate and predictable, they are preferentially optimized. Feedback loops then sustain repeated exposure, reinforcing psychological distress.

This dynamic is particularly visible in eating disorders, like anorexia and bulimia, whose rising incidence among adolescents highlights potential risks of algorithmic misalignment~\citep{radhakrishnan2022pediatric}. We argue that this effect may be rooted in a cognitive heuristic known as \textit{prestige bias}. People tend to preferentially attend to successful or high-status individuals and model their behavior and appearance. In ancestral environments, this heuristic was an efficient strategy for cultural learning---selectively copying successful individuals accelerates the acquisition of adaptive skills and behaviors~\citep{henrich2001evolution}. In algorithmically curated environment of digital media, however, this evolved tendency is systematically exploited~\citep{brady2020attentional}. Platforms expose users to idealized images of bodies, appearances, and lifestyles. Such content elicits envy and is more likely to produce engagement signals and be preferentially recommended. The resulting feedback loops expose users to more representations of ideal bodies, appearances and lifestyles.

Exposure to idealized appearances activates upward social comparison---the tendency to evaluate oneself against others who are perceived as superior on a valued dimension~\citep{choukas2022perfect}. Sustained upward comparison intensifies body dissatisfaction and negative self-evaluation, which are well-established risk factors for depression, anxiety, and eating disorders, particularly among adolescents. Evidence from short-form video platforms suggests that vulnerable users at risk of eating disorders are exposed by recommendation algorithms to more content related to extreme dieting and body image disturbances~\citep{griffiths2024does}.

\subsection{Crowdsourcing and Collective Decision-Making}

Functional misalignment also appears in simpler algorithmic systems that aggregate user behavior without personalization. Crowdsourcing platforms are often framed as examples of the ``wisdom of crowds,'' where aggregating independent judgments helps identify high-quality items~\citep{surowiecki2005wisdom}. However, this logic breaks down under algorithmic mediation.

The MusicLab experiment is a seminal study demonstrating these effects~\citep{salganik2006experimental}, asked participants to evaluate songs by unknown artists. Songs were ranked by prior downloads, which reflects their popularity. Items with early popularity gained visibility through \emph{position bias}~\citep{lerman2014leveraging}, attracting further attention and creating a self-reinforcing feedback loop. This produced two outcomes: \emph{inequality}, where a few songs dominated popularity, and \emph{unpredictability}, where popularity varied widely across identical settings. Importantly, song quality (measured in the control condition) poorly predicted popularity.
Subsequent work showed that cognitive biases and ranking algorithms together can destabilize collective intelligence~\citep{burghardt2020origins}. As cognitive load increases, reliance on heuristics grows, making behavior more predictable and more susceptible to amplification~\citep{burghardt2017myopia}.

Within the functional misalignment framework, these outcomes arise from the same structural mechanisms. Popularity is an imperfect proxy for quality, reflecting an epistemic gap between behavior and value. Feedback loops amplify early variation, producing inequality and instability. Thus, even simple aggregation systems illustrate how optimizing over behavioral signals can lead to collective outcomes misaligned with underlying objectives.


\section{Implications for Human--AI Interaction Research}

The framework developed in this paper has direct implications for how researchers should study and ultimately govern the interactions between humans and algorithmic systems. Prior work often treats outcomes such as polarization, misinformation, and mental health harms as distinct problems amenable to targeted interventions. In contrast, our account shows that these outcomes share structural origins in the coupled dynamics of human behavior and algorithmic optimization. Addressing them therefore requires frameworks that measure, model, and control human--AI systems as dynamical processes. This need is increasingly urgent as AI systems move from passive recommendation toward active participation in social environments. We outline key implications for measurement and experimentation.

\subsection{Measurement Implications}

A central requirement is to distinguish three quantities often conflated in existing work: \emph{behavior} and \emph{preference}, or \emph{value}. Behavioral data (clicks, views, shares) capture what users do, not necessarily what they want. Preferences are internal and context-dependent, and may differ between impulsive System~1 responses and reflective System~2 judgments. Values, commitments, and sense of identity evolve over time and may be reshaped by algorithmic exposure. Treating these as interchangeable risks conflating engagement with welfare.

This distinction has several methodological implications. First, it motivates systematic \emph{preference elicitation}, particularly of reflective (System~2) preferences. In-platform surveys, structured prompts, and deliberative instruments can surface longer-term values that behavioral traces obscure. Second, it calls for \emph{mixed-methods approaches} combining behavioral data with direct measures of wellbeing. Engagement and wellbeing can diverge~\citep{milli2025engagement}, and focusing solely on engagement misses important outcomes. Third, it suggests the need for \emph{multi-objective evaluation}, where engagement is considered alongside preference stability, diversity, and wellbeing. Finally, because preferences may drift  over time, identifying them requires \emph{longitudinal designs} capable of tracking changes in behavior and psychological state rather than relying on cross-sectional snapshots.

\subsection{Experimental Implications}
The limited effectiveness of existing interventions to reduce polarization highlights the need for a structural interventions. Fact-checking, content labeling, cross-partisan exposure, or dialogue have been shown to produce weak, short-lived, or may even backfire to increase polarization instead \citep{bail2018,nyhan2010corrections,holliday2025depolarization,pennycook2019fighting,guess2020exposure}.  By focusing on correcting individual beliefs or behaviors, traditional interventions leave intact the mechanisms that feed division. Reducing polarization therefore requires shifting attention from individual-level interventions to system-level mechanisms---specifically, the amplification dynamics and optimization objectives that make intergroup hostility the most reliable path to engagement in algorithmically curated environments.
Here we discuss experiments that target these underlying mechanisms.

\paragraph{System~1 vs. System~2 Elicitation.}
A first class of experiments seeks to disentangle reactive and reflective signals in user behavior. Engagement conflates impulsive (System~1) and deliberative (System~2) processes, making it an unreliable proxy for user utility. Recent work attempts to separate these signals using temporal dynamics or preference elicitation \citep{agarwal2024system,khambatta2023tailoring}. These approaches suggest that optimizing for engagement may systematically diverge from optimizing for user value, motivating scalable methods for eliciting and integrating reflective preferences.

\paragraph{Prosocial Feed Design.}
A second class of experiments explores alternative ranking objectives. Rather than optimizing solely for engagement, these approaches test whether feeds can be designed to promote outcomes such as reduced polarization, increased diversity, or improved wellbeing. Large-scale field experiments show that modest reductions in polarization can be achieved without substantial engagement losses \citep{stray2026prosocial}. While promising, such interventions remain limited in scope and require systematic exploration across platforms and contexts.

\paragraph{Control of Feedback Dynamics.}
A third class of experiments addresses the control of feedback loops in human--AI systems. Because harms arise from system trajectories rather than isolated decisions, interventions must operate at the level of dynamics. This reframes HAI as a control problem: designing interventions that steer system behavior toward desirable equilibria under uncertainty. Network-based approaches suggest that targeted interventions at key nodes can influence global outcomes \citep{liu2011controllability}. Early evidence indicates that such strategies can reduce harmful behaviors at scale \citep{jahani2026celebrity}, but a general theory of control in these systems remains underdeveloped.

Together, these directions point toward a research agenda that treats human--AI interaction as a coupled dynamical system. Rather than optimizing isolated metrics or deploying one-off interventions, future work should focus on understanding and shaping the mechanisms that generate collective outcomes.

\section{Conclusion}
The framework developed in this paper has direct implications for how algorithmic systems should be designed, measured, governed, and regulated. These implications are worth stating plainly, because the dominant conversations about algorithmic harms in both industry and policy have thus far been organized around assumptions that our paper  calls into question.

The first assumption is that the problem is one of insufficient data. In this view, harms arise because recommendation systems do not know their users well enough, and the remedy is more granular behavioral tracking, longer user histories, and richer feature representations. The framework developed here suggests that this diagnosis is mistaken. The problem is not that algorithms lack data about what users do; it is that what users do is a systematically biased proxy for what users want and need. More data about System~1 behavior does not close the gap between engagement and welfare. Accumulating more of the wrong signal does not produce the right answer.

The second assumption is that the problem is one of insufficient predictive accuracy. This view believes that better models and methods, perhaps reinforcement learning, will learn to serve users more effectively. But functional misalignment is not a prediction error. A system that perfectly predicts which content will maximize engagement and then maximizes it has succeeded at the task it was given while also leading to many unintended negative consequences. The relevant question is not how accurately an algorithm predicts behavior but what it is optimizing for, and whether that objective is appropriately aligned with human welfare. Improving predictive performance within a misaligned objective function does not reduce misalignment---it may accelerate it.

The third assumption is that transparency is sufficient. Algorithmic transparency---disclosure of ranking criteria, platform audits, explainability requirements---is a genuine governance goal that deserves continued support. But transparency alone does not fix misalignment. A fully transparent engagement-maximizing algorithm remains a misaligned one. Users who understand perfectly well that their feed is optimized for engagement, and who are aware of the psychological mechanisms through which it operates, are not protected from those mechanisms. 

This points to the deeper policy implication of the framework. Some values cannot be protected by optimizing more efficiently---they require placing limits on optimizatio itself. Preference stability, epistemic diversity, and emergent unpredictability and disparity are not objectives that can be achieved by adding them as terms in an engagement-maximizing objective function, because they are structural consequences of the feedback loop dynamics in complex systems. Protecting them requires constraints: on the feedback architectures that drive human-algorithm interactions. These are not just technical questions but normative decisions about whose values algorithmic systems should serve and what human capacities they should be prohibited from undermining.

The policy audiences with oversight responsibility for these systems---legislators, regulators, and civil society organizations engaged in technology accountability---face a challenge that is not primarily about detecting bad actors or correcting isolated failures. It is about redesigning the structural incentives that make misalignment the path of least resistance for any platform competing for user attention. That redesign requires a coherent theoretical account of why misalignment arises and how it propagates. The framework developed in this paper is intended as a contribution to that account, and to the research agenda that must accompany it.

\begin{acks}

\end{acks}

\bibliographystyle{ACM-Reference-Format}
\bibliography{references}


\end{document}